\documentclass[sigconf]{acmart}

\usepackage{booktabs} 

\setcopyright{acmlicensed}

\copyrightyear{2018} 
\acmYear{2018} 
\acmDOI{10.1145/ 3195970.3199849}

\acmISBN{978-1-4503-5700-5/18/06}

\acmConference[DAC '18]{DAC '18: The 55th Annual Design Automation Conference 2018}{June 24--29, 2018}{San Francisco, CA, USA}
\acmBooktitle{DAC '18: DAC '18: The 55th Annual Design Automation Conference 2018, June 24--29, 2018, San Francisco, CA, USA}
\copyrightyear{2018}

\acmPrice{15.00}

\usepackage{comment}

\begin{document}
\title[Co-Design of Neural Nets and NN Accelerators]{Co-Design of Deep Neural Nets and Neural Net Accelerators for Embedded Vision Applications}

\author{ Kiseok Kwon,$^{1,2}$ Alon Amid,$^1$ Amir Gholami,$^1$ Bichen Wu,$^1$ Krste Asanovic,$^1$ Kurt Keutzer$^1$}
\affiliation{%
  \institution{$^1$ Berkeley AI Research, University of California, Berkeley\\$^2$ Samsung Research, Samsung Electronics, Seoul, South Korea}}
\email{{kiseok.kwon,alonamid,amirgh,bichen,krste,keutzer}@ berkeley.edu}

\renewcommand{\shortauthors}{Kwon et al.}

\begin{abstract}

Deep Learning is arguably the most rapidly evolving research area in recent
years. As a result it is not surprising that the design of state-of-the-art deep
neural net models proceeds without much consideration of the latest hardware
targets, and the design of neural net accelerators proceeds without much
consideration of the characteristics of the latest deep neural net models.
Nevertheless, in this paper we show that there are significant improvements available if deep neural net models and neural net accelerators are co-designed. This paper is trimmed to 6 pages to meet the conference requirement. A longer version with more detailed discussion will be released afterwards. 

\end{abstract}

%
%



\maketitle

\section{Introduction}
\label{s:intro}

While the architectural design and implementation of accelerators for Artificial Intelligence (AI) is a very popular topic, a more careful review of papers in these areas indicates that both architectures and their circuit implementations are routinely evaluated on AlexNet~\cite{krizhevsky2012imagenet}, a deep neural net (DNN) architecture that has fallen out of use, and whose fat (in model parameters) and shallow (in layers) architecture bears little resemblance to typical DNN models for computer vision. This initial error is compounded by other problems in the  procedures used for evaluation of results. As a result, the utility of many of these NN accelerators on real application workloads is largely unproven. At the same time, contemporary deep neural net (DNN) design principally focuses on accuracy on target benchmarks, with little consideration of speed and even less of energy. Moreover, the implications of DNN design choices on hardware execution are not always understood. 

Thus, a significant gap exists between state of the art NN-accelerator design and state-of-the-art DNN model design. This problem will be carefully reviewed in a longer version of this paper. In this paper we will simply present the results of a coarse-grain co-design approach for closing the gap and demonstrate that
a careful tuning of the accelerator architecture to a DNN model can lead to a $1.9-6.3\times$ improvement in speed in running that model. 
We also show that integrating hardware considerations into the design of a neural net model can yield an improvement of $2.6\times$ in speed and $2.25\times$ in energy as compared to SqueezeNet~\cite{iandola2016squeezenet} ($8.3\times$ and $7.5\times$ compared to AlexNet), while improving the accuracy of the model. 

The remainder of this paper is broadly organized as follows. In Section \ref{s:app_constraints}, we begin with a brief introduction to applications in embedded computer vision, and their natural constraints in speed, power, and energy. In Section  \ref{s:design_of_nn}, we discuss the design of NN accelerators for these embedded vision applications. In Section \ref{s:codesign}, we turn our focus to the co-design of DNN and NN accelerators. We end with our conclusions.
\section{Computer Vision Applications and Their Constraints}
\label{s:app_constraints}

The precise implementation constraints for an embedded computer vision application can vary widely, even for a single application area such as autonomous driving. In this paper we are particularly concerned with the design problems for computer vision applications that run on in a limited form-factor, on battery power, and with no special support for heat dissipation, but nevertheless have real-time latency constraints. Altogether, these form-factor and packaging constraints imply limits on power and memory. Optimizing for battery life naturally constrains the energy allotted for the computation. We further presume that overriding these concerns, the application has fixed accuracy requirements (such as classification accuracy) and latency requirements. Thus, an embedded vision application must guarantees a level of accuracy, operate within real-time constraints, and optimize for power, energy, and memory footprint.

For all the variety of computer vision applications described earlier in this section, there are a few basic primitives of kernels out of which these applications are built. For perception tasks where the goal is to understand the environment, the most common tasks include: image classification, object detection, and semantic segmentation.

Image classification aims at assigning an input image one label from a fixed set of categories. A typical DNN model takes an image as input and compute a fixed-length vector as output. Each element of the output vector encodes a probability score of a certain category. Depending on specific dataset, typical input resolutions to a DNN can vary from $32\times32$ to $227\times227$. Normally image classification is not sensitive to spatial details. Therefore, several down-sampling layers are adopted in the network to reduce the feature map's resolution until the output becomes a single vector for representing the whole image. 

Object detection and semantic segmentation are more sensitive to image resolutions~\cite{ashraf2016shallow,wu2017squeezedet}. Their input size can range from hundreds to thousands of pixels, and the intermediate feature map usually cannot be over sub-sampled in order to preserve spatial details. As a result, DNN for object detection and semantic segmentation have much larger memory footprint. As image classifiers form the trunk of other DNN models, we will focus on image classification in the remainder of the paper.
\section{Design of NN Accelerators for Embedded Vision}
\label{s:design_of_nn}
The power, energy, and speed constraints for embedded vision applications
discussed in the previous section naturally motivate a specialized accelerator
for the inference problem of NNs. 
The typical approach to micro-architectural design of accelerators is to find a representative workload, extract characteristics, and tailor the micro-architecture to that workload
\cite{hennessy2011computer}. However, as DNN models are evolving quickly we feel that co-design of DNN models and NN accelerators is especially well motivated.  

\begin{figure*}[!htbp]
    \centering
    \includegraphics[width=0.85\textwidth]{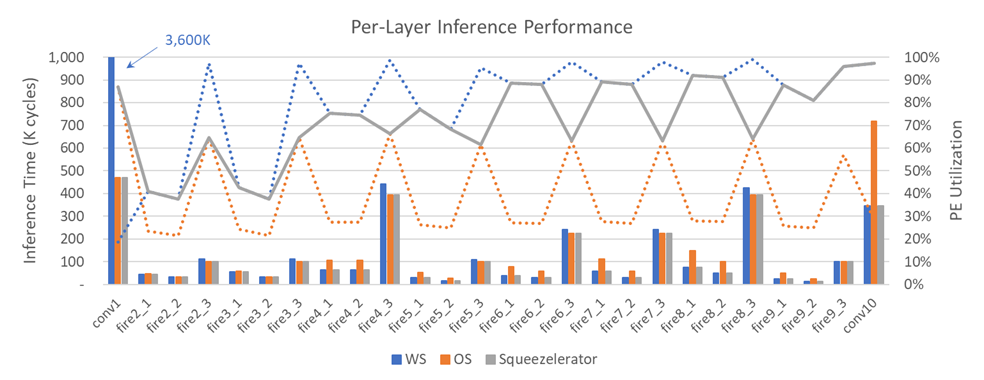}
    \caption{
     Per-layer inference time (bar) and utilization efficiency (dotted and solid lines) of SqueezeNet v1.0 on the reference WS/OS architectures and Squeezelerator.
    }
    \label{fig:squeezenet_per_layer}
\end{figure*}

\subsection{Key Elements of NN Accelerators} 
Spatial architecture (\textit{e.g.}~\cite{chen2016eyeriss}) are a class of accelerator architectures that exploit the high
computational parallelism using direct communication between an array of relatively
simple processing elements (PEs). Compared to SIMD architectures,
spatial architectures have relatively low on-chip memory bandwidth per PE, but they
have good scalability in terms of routing resources and memory bandwidth. 
Convolutions constitute 90\% or more of the computation in DNNs for embedded vision, and are therefore called convolutional neural netowrks (CNN). 
Thanks
to the high degree of parallelism and data reusability of the convolution, the
spatial architecture is a popular option for accelerating these CNN/DNNs \cite{chen2016eyeriss,jouppi2017datacenter,du2015shidiannao,moons201714,lu2017flexflow}. Hereafter,
we restrict the type of the NN accelerators we consider  to  spatial architectures.

In order to exploit the massive parallelism, NN accelerators contain a large
number of PEs that run in parallel. A typical PE consists of a MAC unit and a
small buffer or register file for local data storage. Many accelerators employ
a two-dimensional array of PEs, ranging in size from as small as $8\times8$~\cite{du2015shidiannao} to as large as $256\times256$~\cite{jouppi2017datacenter}. However, an increase in the number of PEs
requires an increase in the memory bandwidth. A MAC operation has three input
operands and one output operand, and supplying these operands to hundreds of
PEs using only DRAM is limited in terms of bandwidth and energy consumption.
Thus, NN accelerators provide several levels of memory hierarchy to provide
data to the MAC unit of the PE, and each level is designed to take advantage of
the data reuse of the convolutional layer to minimize access to the upper
level. This includes global buffers (on-chip SRAMs) ranging from tens of KBs to
tens of MBs, interconnections between PEs, and local register files in the PE.
The memory hierarchy and the data reuse scheme are one of the most important
features that distinguish NN accelerators. Some accelerators also have
dedicated blocks to process NN layers other than convolutional layers \cite{du2015shidiannao,moons201714,lu2017flexflow}. Since
these layers have a very small computational complexity, they are usually
processed in a 1D SIMD manner.

\subsection{A Taxonomy of NN Accelerator Architectures}

There are several features that distinguish NN accelerators, and the following are some examples.

\begin{itemize}
  \item PE: data format 
  (log, linear,
  floating-point), bit width, implementation of arithmetic unit (bit-parallel, bit-serial~\cite{judd2016stripes}), 
  data to reuse (input, weight, partial sum)
  \item PE array: size, interconnection topology, data reuse, algorithm mapping
  \item global buffer: configuration (unified~\cite{chen2016eyeriss}, dedicated~\cite{jouppi2017datacenter}), memory type (SRAM, eDRAM~\cite{chen2014dadiannao})
  \item data compression, sparsity exploitation~\cite{han2016eie,parashar2017scnn}, multi-core configuration
\end{itemize}

Eyeriss~\cite{chen2016eyeriss} proposed a useful taxonomy that classifies NN
accelerators according to the type of data each PE locally reuses. Since the
degree of data reuse increases as the memory hierarchy goes down, this type of
classification shows the characteristic reuse scheme of NN accelerators. Among
the four dataflows, weight stationary (WS), output stationary (OS), row
stationary (RS), and no local reuse (NLR), two are introduced here.

\paragraph{Weight Stationary} The weight stationary (WS) dataflow is designed
to minimize the required bandwidth and the energy consumption of reading model 
weights by maximizing the accesses of the weights from the register file at the PE.
The execution process is as follows. The PE preloads a weight of the
convolution filters to its register. Then, it performs MAC operations over the
whole input feature map. The result of the MAC is sent out of the PE in each
cycle. Afterwards, it moves to the next element and so forth.

There are several ways to map the computation to multiple PEs. One example is
to map the weight matrix between the input and output channels
to the PE array. 
Such hardware takes the form of a general
matrix-vector multiplier. TPU~\cite{jouppi2017datacenter} has a $256\times256$ PE array, which performs matrix-vector multiplications over a stream of
input vectors in a systolic way. 
The input vectors are passed to each
column in the horizontal direction, and the partial sums of PEs are propagated
and accumulated in the vertical direction. 
In this way, TPU can also reuse
inputs up to 256 times and reduce partial sums up to 256 times at the PE array
level.

\paragraph{Output Stationary} The output stationary (OS) dataflow is designed
to maximize the accesses of the partial sums within the PE. In each cycle, the
PE computes parts of the convolution that will contribute to one output pixel,
and accumulates the results. Once all the computations for that pixel are
finished, the final result is sent out of the PE and the PE moves to work on a
new pixel.

One example of the OS dataflow architecture is
ShiDianNao~\cite{du2015shidiannao}, which maps a 2D block of the output feature
map to the PE array. It has an 8x8 PE array, and each PE handles the processing
of different activations on the same output feature map. The PE array performs
$F_x \times F_y$ filtering on a  $(F_x+7) \times (F_y+7)$ block of the input
feature map over $F_x \times F_y$ cycles. In the first cycle, the top left $8\times8$
pixels of the input block is loaded into the PE array. In the following cycles,
most of the input pixels are reused via mesh-like inter-PE connections, and
only small part of the input block is read from the global buffer. The
corresponding weight is broadcasted to all PEs every cycle.

\section{Co-design of DNNs and NN accelerators}
\label{s:codesign}

In this section we describe the co-design of DNNs and NN accelerators.
Because the design of either a DNN or a NN accelerator is a significant enterprise, the co-design of these is necessarily a coarse grained process.
Thus, we first describe the design of the Squeezelerator, a NN accelerator intended to accelerate SqueezeNet. We then continue with a discussion of the design of SqueezeNext, a DNN designed with the principles described in 
\cite{iandola2017small} and
particularly tailored to execute efficiently on the Squeezelerator. Finally, we discuss the additional tune-ups of the Squeezelerator for SqueezeNext.
There are many design considerations in this process that are given in more detail in a longer version of this paper. 

\subsection{Tailoring the design of a NN accelerator to a DNN}
The accelerator, Squeezelerator, was designed to accelerate  SqueezeNet v1.0 and to be used as an IP block in a systems-on-a-chip (SOC) targeted for mobile or IoT applications. According to our simulations the accelerator also shows good performance for a variety of neural network architectures such as MobileNet.

\subsubsection{ Characteristics of the target DNN architecture}
\label{ss:characteristics}

\begin{table}[!htbp]
\caption {
  Relative percentage of MAC operations/total operations for each layer type in each of the DNN Networks
}
\label{t:proportion}
\centering
\begin{tabular}{l|r|r|r|r}
\midrule
  Network           & Conv1  & $1 \times 1$ & $F \times F$ & DW \\
\midrule
  AlexNet           &  20\%  &   0\%  & 69\%  & 0\%  \\
  1.0 MobileNet-224 &   1\%  &  95\%  &  0\%  & 3\%  \\
  Tiny Darknet      &   5\%  &  13\%  & 82\%  & 0\%  \\
\midrule
  SqueezeNet v1.0   &  21\%  &  25\%  & 54\%  & 0\%  \\
  SqueezeNet v1.1   &   6\%  &  40\%  & 54\%  & 0\%  \\
  SqueezeNext       &  16\%  &  44\%  & 40\%  & 0\%  \\
\bottomrule
\end{tabular}
\end{table}

Based on the analysis of previous experimental results, we classify  convolution layers into four categories (see Table~\ref{t:proportion}): the first convolutional layer, pointwise convolution (\textit{i.e.} $1\times1$), $F\times F$ convolution (where $F>1$), and depthwise convolution (DW). 
The following numbers are from simulations on a 32x32 PE Squeezelerator. Depending on the size of the feature map and the number of channels, our simulations indicate that $1\times1$ convolutional layers are 1.4x to 7.0x faster on a WS dataflow architecture than on a OS dataflow. 
In contrast, relative to the WS dataflow architecture, the first convolutional layer is 1.6x to 6.3x faster on the  OS dataflow architecture and the depthwise convolutional layers are 19x to 96x faster on the OS dataflow architecture. 
In the case of the normal $3\times3$ convolutions, various factors affect actual acceleration speed of the OS dataflow including the size of the feature map and the sparsity of the filters. Therefore, each layer configuration must be simulated to determine which architecture is best. 
As the DNN inference computation is  statically schedulable,  simulation results can be used to determine the dataflow approach (WS or OS) that best executes the $3\times3$ convolution. 
Table~\ref{t:proportion} shows the relative percentage of computation devoted to each layer type in a variety of DNNs. There is a large variation in the percentages for each category over these DNN models, and as a result the proportion of the layer operations which are well-suited to the WS dataflow ranges from 0\% to 95\%. While initially focused on supporting SqueezeNet, this layer analysis led to the key design principal of the Squeezelerator: to achieve high efficiency for the entire DNN model, the accelerator architecture must be able to choose WS dataflow or OS on a layer by layer basis.

\begin{figure}[!htbp]
    \centering
    \includegraphics[width=0.45\textwidth]{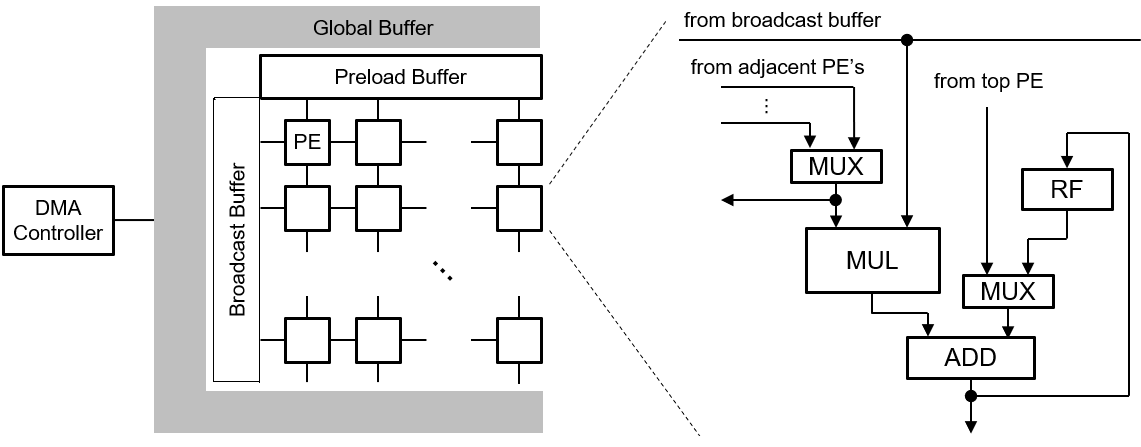}
    \caption{The block diagram of Squeezelerator (left) and PE (right)}
    \label{fig:squeezelerator}
\end{figure}

Thus, the design of the Squeezelerator is based on the layer by layer simulation as described above. 
As shown in Figure~\ref{fig:squeezelerator}, it consists of a PE array, a global buffer, a preload buffer, a stream buffer, and a DMA controller. Intended for SOC deployment, the PE array consists of $N\times N$ PEs (for $N$=8 to 32) and inter-PE connections to handle the convolution and FC layer operations. Each PE is connected to adjacent PEs in a mesh structure, as well as to the broadcast buffer. The PEs located at the top and the bottom row of the mesh are additionally connected to the preload buffer and the global buffer, respectively. (This communication topology is appropriate for a SOC, but more limited than GPU designs. As a result, layer execution times will differ on GPUs as well.) The preload buffer prepares the data to be transferred to the PE array before the operation starts, and the stream buffer prepares the data to be continuously transferred to the PE array during the operation. The global buffer consists of 128KB on-chip SRAM and switching logic. It is connected with preload buffer, stream buffer, and DMA controller. A PE contains a MUX for selecting one of several input sources, a 16-bit integer multiplier, an adder for accumulating the multiplication result, and a register file for storing the intermediate result of the computation. In order to support two dataflows, we implemented all the interconnections and functions required for both dataflows. The area overhead is minimized by providing different data to the PE array in each mode. For example, the broadcast buffer provides the input activations in the WS mode, while it provides the weights in the OS mode.

\subsubsection{Operation sequence}

\begin{figure*}[!htbp]
    \centering
    \includegraphics[width=0.89\textwidth]{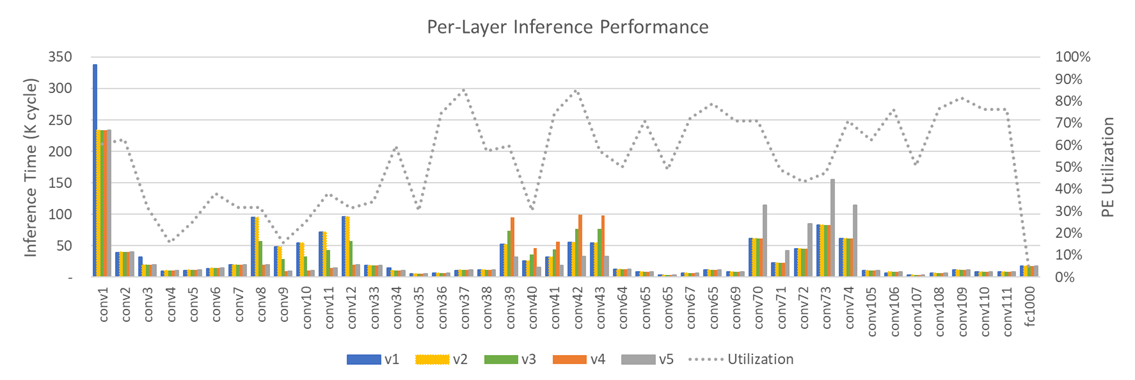}
    \caption{
     Per-layer inference time (lower is better) is shown along the left y-axis for five variants (v1-v5) of 1.0-SqNxt-23 architecture. PE utilization is shown by the dotted line and the right y-axis. As one can see, the initial layers have very low utilization which adversely affects inference time and energy consumption.
    }
    \label{fig:sqnxt_per_layer}
\end{figure*}

Squeezelerator processes the DNN layer by layer, and it can be configured to select the dataflow style (OS or WS) for each layer, and no overhead is incurred by switching between dataflow styles. While the accelerator is running in the OS dataflow mode, each PE is responsible for different pixels in the 2D block of the output feature map. Every cycle the corresponding input and weight are supplied to each PE through inter-PE connection and from the broadcast buffer, respectively. The operation sequence is as follows. First, a 2D block of the input feature map is preloaded into the PE array from the preload buffer. Then, the stream buffer broadcasts a weight to all the PEs, and each PE multiplies the input by the weight and accumulates the result in the local register file. For a $N \times N$ filter, this step is repeated $N^2$ times with different input and weight data. Instead of reading the input from the preload buffer every time, each PE receives the data from one of the neighboring PEs. The whole process is repeated with different input channels. When the computation for the output block is finished, the result of each PE is stored to the global buffer. This final step takes additional processing time. To reduce the energy consumed by the global buffer, PEs reuse each input they receive across different filters. In addition, the stream buffer broadcasts only non-zero weights to reduce the execution time by skipping unnecessary computations.

In the WS dataflow mode, a PE row and a PE column correspond to one input channel and one output channel, respectively. In this way, each PE is responsible for different elements of the weight matrix. Contrary to the OS mode, the $16\times16$ ``weights'' are preloaded into the PE array. Then, the stream buffer broadcasts pixels from 16 different ``input channels'' to the PE array, and each PE multiplies the input by its own weight. Each PE column sums the multiplication results by forming a chain of adders from the top PE to the bottom PE. This process is repeated until all the pixels in the input feature maps are accessed. 

\vspace*{-5mm}
\subsubsection{Experimental results}

A performance estimator evaluates the execution cycle and the energy consumption of Squeezelerator. Results describe inference times of individual images (\textit{i.e.} batch size = 1) from the ImageNet benchmark suite \cite{deng2009imagenet}. A batch size of one gives less opportunity for data reuse, but reflects typical usage in embedded vision applications for mobile phones or automotive perception.
The time consumed by the PE array and the buffers reflects the micro-architecture, and the DRAM access time is approximated by using two numbers: latency and effective bandwidth. The numbers used in the experiments are 100 cycles and 16 GB/s, respectively. In order to hide the data transfer time between the DRAM and the global buffer, we used double buffering~\cite{kim2001data}. If the memory footprint of the layer exceeds the capacity of the buffer, some of the six convolution loops are tiled. The size of the tile and the order of loops that give the shortest execution time are selected. We followed the methodology used by \cite{chen2016eyeriss} for energy estimation. It calculates the number of accesses of the MAC units and each memory layer, and then multiplies each by its unit energy, which is normalized by the energy consumption of the MAC unit. Here we modified the unit energy slightly to match this hardware configuration. During simulation we conservatively model the sparsity, \textit{i.e.} the number of zero weights, of each DNN layer at 40\%.


We first evaluate Squeezelerator with the target DNN, SqueezeNet v1.0. Figure~\ref{fig:squeezenet_per_layer} shows the inference time and utilization per layer 
for the reference and the proposed architectures.
The overall trend is similar to that of the WS architecture, but the performance of the first layer is noticeably improved. For most of the $3 \times 3$ convolutions, the accelerator chooses OS dataflow. Although the OS dataflow exploits the sparsity of the filters, it does not show a significant improvement over the WS. In the early layers, it is due to the additional time for transferring the computation results to the global buffer. In the latter layers, the mismatch between the size of the PE array and the size of the feature map is the main cause of the performance degradation. Comparing the total processing time, the proposed structure shows performance improvement of 26\% and 106\% compared to the reference OS and WS architectures, respectively.

\begin{table}[!htbp]
\caption {
  Speed and Energy Improvements of Squeezelerator over OS or WS architectures
}
\label{t:estimation_results}
\centering
\begin{tabular}{l|r|r|r|r}
\midrule
  Network               & \multicolumn{2}{c|}{Speedup} & \multicolumn{2}{c}{Energy reduction} \\
                        & vs OS & vs WS & vs OS & vs WS \\
\midrule
  AlexNet           & $1.00\times$  & $1.19\times$  & -2\%  &  6\%  \\
  1.0 MobileNet-224 & $1.91\times$  & $6.35\times$  &  8\%  &  6\%  \\
  Tiny Darknet      & $1.14\times$  & $1.32\times$  &  0\%  & 24\%  \\
\midrule
  SqueezeNet v1.0   & $1.26\times$  & $2.06\times$  &  6\%  & 23\%  \\
  SqueezeNet v1.1   & $1.34\times$  & $1.18\times$  &  8\%  & 10\%  \\
  SqueezeNext       & $1.26\times$  & $2.44\times$  &  0\%  & 20\%  \\
\bottomrule
\end{tabular}
\end{table}

Table~\ref{t:estimation_results} shows the performance improvement of the Squeezelerator over the reference architectures on AlexNet (just for comparison) and a variety of lightweight DNNs. The improvement over the OS architecture has a high correlation with the proportion of the $1\times1$ convolutions in the network. The benefits of supporting two dataflow architectural styles are obvious in the case of MobileNet. 
Because a naive WS architecture does not efficiently accelerate the depthwise convolutions, these layers occupy much larger execution time than the pointwise convolutional layers, even though they account for only 3\% of the total number of computations. On the other hand, the $1\times1$ convolutions, which account for 95\% of the total computation, greatly reduces the acceleration performance of the OS architecture. The proposed architecture achieves about 2x and 6x speed up compared to the OS and WS architectures.

While space does not permit it here, a more detailed per-layer evaluation will be given for each DNN model in a longer version of this paper. AlexNet shows the least performance improvement because it takes up 80\% of energy and 73\% of its run time in the three fully-connected layers, which cannot take advantage of hardware acceleration by either dataflow architecture. MobileNet shows small savings on the energy consumption relative to its significant performance improvement, because  DRAM access consumes a larger proportion of total energy consumption in this network than in other DNNs. This is related with the low data reusability of the pointwise convolutions and the depthwise convolutions. The energy reduction of SqueezeNet V1.0 and Tiny Darknet is due to their larger proportion of layers that are suited to the OS dataflow.

\subsection{ Co-Design of a DNN and NN accelerator}

\begin{figure}[!htbp]
    \centering
    \includegraphics[width=0.35\textwidth]{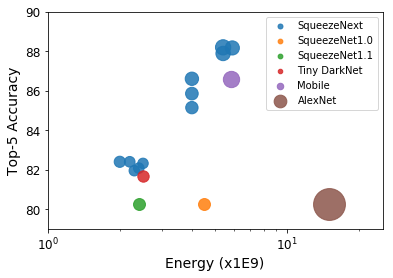}\\
    \includegraphics[width=0.35\textwidth]{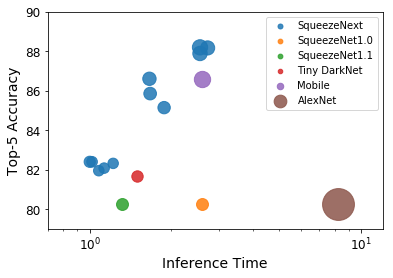}
    \caption{The spectrum of accuracy versus energy and inference speed for SqueezeNext, SqueezeNet (v1.0 and v1.1), Tiny DarkNet, and MobileNet. SqueezeNext shows superior performance (in both plots higher and to the left is better).
}
    \label{fig:sqnxt_accuracy_energy}
\end{figure}

Earlier in this section, we presented the design of the Squeezelerator NN accelerator tailored for SqueezeNet and applied to other small DNN models. Now we continue with our co-design process and describe SqueezeNext ~\cite{squeezenext}, a new family of neural networks for embedded systems, which was designed by performing detailed analysis with an architectural simulator for Squeezelerator .

SqueezeNext was designed by studying previous DNN models such as SqueezeNet with the aim of using structure of layers to further reduce model parameters, and avoiding Mobilenet's depthwise separable convolutions that have poor Arithmetic Intensity (Ops/MAC per byte of memory accessed).
%

Studying the hardware utilization of different layers of SqueezeNext on the Squeezelerator revealed that initial layers had low MACs/cycle counts, which noticeably affected hardware performance, as shown in Figure~\ref{fig:sqnxt_per_layer}.
One important optimization is filter size reduction for the first layer from $7\times7$ to $5\times5$; this layer has significant impact on inference time as its input feature map is relatively large.
Another contributing factor to poor hardware utilization is the small number of channels in the initial layers. 
In this situation not all PEs will be utilized, and because there is limited data reusability there is limited opportunity to hide memory latency. 
One solution to this would be to simply reduce the number of layers early in the DNN; however, a naive reduction may lead to a degradation in accuracy.
Instead, we reduce the number of layers early in the DNN and assign more layers to later stages of it that have higher hardware utilization. 
While this simple change results in a very small change in the overall  MACs used in inference, it reduces  both energy and inference time~\cite{squeezenext}.  
Five different variants of these two classes of optimizations are shown in Figure~\ref{fig:sqnxt_per_layer}.   
In fact, the optimized versions have slightly better accuracy as compared to the initial variant. Following this design of SqueezeNext we returned to the co-design of the Squeezelerator and fine-tuned the hardware utilization by doubling the register file size from 8 to 16. The combination of these optimizations resulted in SqueezeNext being $2.59\times$ faster and $2.25\times$ more energy efficient than SqueezeNet 1.0  (and $8.26\times$ and $7.5\times$ when compared to AlexNet), without any degradation in accuracy.

Figure~\ref{fig:sqnxt_accuracy_energy} shows the spectrum of accuracy vs power and accuracy vs inference time
for different DNN families. Ideally, we would like higher accuracy with smaller power and inference time.
As we can see SqueezeNext family provides such favorable solutions which allows the user to select the right DNN from this family based
on the target application's constraints.

\section{Conclusions}
\label{s:conclusions}

Embedded vision applications bring power, energy, memory, and speed constraints. In this paper we have illustrated a coarse grain co-design approach for the design of DNNs and NN accelerators that meet these constraints. Our efforts at a NN accelerator led to the novel design of the Squeezelerator, which can perform either  weight-stationary dataflow or a output-stationary dataflow on a layer-by-layer basis, with no additional latency. on popular DNNs for mobile applications this accelerator design  is $1.1\times$ - $6.35\times$ faster than accelerators that use only a single dataflow architecture. 
To illustrate the additional value of tailored DNN design to the accelerator, we revisited the design of SqueezeNet and produced the SqueezeNext family. Some members of the SqueezeNext family are $2.26\times$ faster than SqueezeNet 1.0, improve the energy by $2.25\times$, and are more accurate on image classification benchmarks (we achieve 59.2\% top-1 vs 57.1\% of SqueezeNet)~\cite{squeezenext}. 
We completed our design study by then revisiting the design of the Squeezelerator running SqueezeNext. As SqueezeNext has similar layer characteristics to SqueezeNet, only some fine-tuning of register file size was required to optimize local data reuse. We present a more detailed description of our design choices and our evaluation methodology in an extended version of this paper, but we hope to have demonstrated the value of co-design through this study.

\bibliographystyle{ACM-Reference-Format}
\bibliography{acmart}

\end{document}